\pdfoutput=1

\documentclass[11pt]{article}
\usepackage{epsf,amsmath,amssymb,graphicx,scalefnt,rotating,enumitem,fancyhdr}
\numberwithin{equation}{section}
\usepackage[acronym]{glossaries}
\let\oldglsentryshort\glsentryshort
\renewcommand{\glsentryshort}[1]{\abbrev{\oldglsentryshort{#1}}}

\usepackage[final]{showkeys}
\usepackage[noadjust]{cite}
\usepackage{filemod}
\usepackage{framed}
\usepackage{comment}
\usepackage[pdftex, colorlinks=true, linkcolor=black, filecolor=black,
  citecolor=black, urlcolor=black, draft=false, bookmarks,
  bookmarksnumbered=true, plainpages=false,
  linktocpage={true}]{hyperref}
\usepackage[affil-it]{authblk}
\usepackage[usenames,dvipsnames]{color}
\usepackage{tikzsymbols}
\usepackage[capitalize]{cleveref}

\usepackage{slashed}
\usepackage{dsfont}
\usepackage{braket}

\newcommand{\lrD}{\overset\leftrightarrow{D}}
\newcommand{\lD}{\overset\leftarrow{D}}
\newcommand{\be}{\begin{equation}}
\newcommand{\ee}{\end{equation}}
\def\qbar{\bar{q}}
\def\chibar{\overline{\chi}}
\def\wmcO{\widehat{\mcO}}
\def\mcO{{\mathcal O}}

\newif\ifshownewacro
\includecomment{private}
\shownewacrotrue  

\newcommand{\two}{two}
\newcommand{\three}{three}
\newcommand{\four}{four}

\newcommand{\eqindentspace}{em}
\newcommand{\eqindent}[1]{\hspace{#1\eqindentspace}}
\newcommand{\lmut}{L_{\mu t}}
\newcommand{\ccf}{C_\text{F}}
\newcommand{\ctr}{T_\text{R}}
\newcommand{\nc}{n_\text{c}}

\newcommand{\nf}{n_\text{f}}
\newcommand{\cca}{C_\text{A}}

\newcommand{\bare}{\text{\abbrev{B}}}

\newcommand{\citere}[1]{Ref.\,\cite{#1}}
\newcommand{\citeres}[1]{Refs.\,\cite{#1}}
\newcommand{\code}[1]{\textsl{#1}}
\newcommand{\abbrev}[1]{{\scalefont{.9}#1}}
\newcommand{\EulerGamma}{\gamma_\text{E}}
\newcommand{\ep}{\epsilon}
\newcommand{\api}{a_\text{s}}
\newcommand{\dd}{\mathrm{d}}
\newcommand{\deriv}[3]{\frac{\partial\ifthenelse{\equal{#1}{}}{}{^{#1}}%
    #2}{\partial #3\ifthenelse{\equal{#1}{}}{}{^{#1}}}}
\newcommand{\dderiv}[3]{\frac{\dd\ifthenelse{\equal{#1}{}}{}{^{#1}}%
    #2}{\dd #3\ifthenelse{\equal{#1}{}}{}{^{#1}}}}

\newcommand{\msbar}{\ensuremath{\overline{\text{\abbrev{MS}}}}}

\newcommand{\lhs}{l.h.s.}
\newcommand{\rhs}{r.h.s.}
\newcommand{\wrt}{w.r.t.}

\newcommand{\myacrodef}[3]{\newacronym[shortplural=\abbrev{#2}s]{#2}{#2}{#3}\newcommand{#1}{\gls{#2}}}

\newacronym[\glslongpluralkey={effective field theories},%
  shortplural=\abbrev{EFT}s]{EFT}{EFT}{effective field theory}

\newacronym[\glslongpluralkey={generalized parton densities},%
  shortplural=\abbrev{GPD}s]{GPD}%
{GPD}{generalized parton density}

\newcommand{\gpds}{\glspl{GPD}}

\myacrodef{\dglap}{DGLAP}{Dokshitzer-Gribov-Lipatov-Altarelli-Parisi}
\myacrodef{\sftx}{SFTX}{short-flow-time expansion}
\myacrodef{\rg}{RG}{renormalization group}
\myacrodef{\gf}{GF}{gradient flow}
\myacrodef{\gff}{GFF}{gradient-flow formalism}
\newcommand{\qcd}{\abbrev{QCD}}

\myacrodef{\lhc}{LHC}{Large Hadron Collider}
\myacrodef{\uv}{UV}{ultra-violet}
\myacrodef{\lo}{LO}{leading order}
\myacrodef{\nlo}{NLO}{next-to-leading order}
\myacrodef{\nnlo}{NNLO}{next-to-next-to-leading order}
\myacrodef{\pdf}{PDF}{parton distribution function}
\newcommand{\pdfs}{\glspl{PDF}}
\myacrodef{\sm}{SM}{Standard Model}
\myacrodef{\bsm}{BSM}{beyond-the-Standard Model}
\allowdisplaybreaks
\newcommand{\RHheaderline}{\textsf{TTK-25-41, P3H-25-094, November 2025}}
\fancypagestyle{firstpage}
{
  
  \fancyhead[R]{\RHheaderline}
}
\title{Short-flow-time expansion of non-singlet twist-\two\ operators
at next-to-next-to-leading order QCD}

\author[1]{Robert V. Harlander}
\author[1]{Jonas T. Kohnen}
\author[1,2,3]{Andrea Shindler}
\affil[1]{TTK, RWTH Aachen University, 52056 Aachen, Germany}
\affil[2]{Nuclear Science Division, Lawrence Berkeley National Laboratory, Berkeley, CA 94720 USA}
\affil[3]{Department of Physics, University of California, Berkeley, CA 94720, USA}

\date{}
\glsresetall 

\begin{document}
\maketitle
\thispagestyle{firstpage}
\begin{abstract}
  The gradient-flow formalism provides a framework for the direct
  determination of moments of \pdfs\ from lattice \qcd\ calculations.  Their
  conversion from the gradient-flow scheme to \msbar\ requires the matching
  coefficients of the short-flow-time expansion, which can be computed
  perturbatively.  We determine these coefficients for the first six
  non-singlet \pdf\ moments up to next-to-next-to-leading order in the strong
  coupling.
\end{abstract}
\parskip.0cm
\tableofcontents
\parskip.2cm
\glsresetall 



\section{Introduction}
\label{sec:intro}

Theoretical cross-section predictions for hadronic scattering processes rely
on the factorization~\cite{Collins:1989gx} into a perturbatively accessible
partonic subprocess and non-perturbative contributions associated with the
initial- and final-state hadrons.  Those related to the initial state are
encoded in the \pdfs, which describe the probability of finding a parton
carrying a fraction~$x$ of the hadron momentum.  Their dependence on the
factorization scale $Q$ is governed by the \dglap{} evolution
equations~\cite{Gribov:1972ri,Altarelli:1977zs,Dokshitzer:1977sg}, whose
kernels are the splitting functions.

A quantitative determination of the \pdfs\ requires both the perturbative
splitting functions and suitable non-perturbative initial conditions, i.e.\
the $x$ dependence of the parton densities at a reference scale~$Q_0$.  The
splitting functions are calculable in perturbation theory and are known up
to \three-loop order~\cite{Moch:2004pa,Vogt:2004mw}, with partial \four-loop
results now
available~\cite{Gehrmann:2023cqm,Falcioni:2023tzp,Gehrmann:2023iah,
Falcioni:2024qpd,Falcioni:2024xyt,Kniehl:2025jfs,Kniehl:2025ttz}.  The initial
conditions, on the other hand, are typically obtained by fitting
phenomenological parametrizations of \pdfs\ to data, although significant
progress has recently been made toward their direct computation using lattice
\qcd.

A first-principles determination of \pdfs\ is indeed highly desirable.  It
would clarify the non-perturbative structure of \qcd, provide systematically
improvable input for collider phenomenology, and supply theoretical benchmarks
that are independent of global fits and could be combined with them to
increase precision.  Such progress would strengthen precision tests of
the \sm\ and improve the sensitivity to possible signals of physics \bsm\ at
present and future
facilities~\cite{Amoroso:2022eow,Abir:2023fpo,LHeC:2020van}.

Lattice \qcd\ offers a natural framework for such a determination.  While
lattice calculations have achieved remarkable precision in many areas of
hadronic physics, obtaining phenomenologically relevant results for \pdfs\
remains challenging. The main reason for this is that \pdfs\ are defined
through light-cone correlations, whereas lattice gauge theory is formulated in
Euclidean space.  Several theoretical frameworks, such as quasi-\pdfs,
pseudo-\pdfs, and related approaches provide access to the $x$-dependence
of \pdfs\ from Euclidean
correlators~\cite{Liu:1993cv,Aglietti:1998mz,Detmold:2005gg,Ji:2013dva,Radyushkin:2017cyf,Orginos:2017kos,Ma:2014jla,Braun:2007wv,Chambers:2017dov,Gao:2023lny}. 
These methods establish a systematic connection between Euclidean observables
and light-cone parton physics, although they require careful control of
short-distance effects and hadron momenta to ensure reliable access to the
light-cone regime.

An alternative and long-established approach focuses on the Mellin moments
$\langle x^{n-1}\rangle$ ($n\in\mathds{N}$) of the \pdfs, which can be
expressed as matrix elements of local twist-\two\ operators accessible on the
lattice.  However, the reduced symmetry of discretized space-time compared to
the continuum leads to mixing of these operators with lower-dimensional ones,
resulting in divergences that scale as inverse powers of the lattice
spacing~$a$~\cite{Kronfeld:1984zv}.  This limits the attainable precision for
higher moments, and conventional lattice determinations beyond the lowest
few~\cite{Martinelli:1987zd,Martinelli:1987bh,Gockeler:1996mu,
Alexandrou:2020gxs,Alexandrou:2021mmi} remain impractical.

Recently, it was shown~\cite{Shindler:2023xpd} that these power divergences
can be avoided entirely by employing the \gff~\cite{Narayanan:2006rf,
Luscher:2010iy,Luscher:2011bx,Luscher:2013cpa}.  In this approach,
high-momentum modes of the flowed fields are exponentially suppressed by the
flow time~$t$, rendering matrix elements of composite operators \uv-finite and
admitting a well-defined continuum limit.  Since the \gff\ preserves the full
rotational symmetry of Euclidean space, operators can be organized into
irreducible representations of O($4$). Mixing with lower-dimensional operators
is absent for $t>0$, and so are the associated power divergences. Building on
this property, \citere{Shindler:2023xpd} proposed a new approach that employs
flowed twist-\two\ operators to determine higher Mellin moments of the \pdfs.

To relate the flowed matrix elements at finite flow time~$t>0$ to the standard
renormalization scheme used in phenomenology, one must convert them to the
\msbar\ scheme defined at $t=0$.  This is achieved through a \sftx\ \cite{Luscher:2011bx,Suzuki:2013gza,Luscher:2013vga}, in which
the flowed twist-\two\ operators are expressed as a series of local operators
renormalized in the \msbar\ scheme and multiplied by perturbatively calculable
matching coefficients.  At \nlo\ in \qcd, the matching coefficients for the
moments $\langle x^{n-1}\rangle$ have been computed for general
$n\in\mathds{N}$~\cite{Shindler:2023xpd,Shindler:2024grr}.  From other
applications of the \sftx, it is known that higher-order corrections to the
matching coefficients are important in several
respects~\cite{Iritani:2018idk,Harlander:2022vgf,Black:2024iwb}. First and
foremost, as in any perturbative calculation, they are crucial for reducing
the systematic uncertainties from missing higher orders, of course. But in the
context of the \sftx, they are also found to stabilize the extrapolation to
$t\to 0$.

For this reason, in the present work we extend the calculation of the matching
coefficients for the \pdfs\ to \nnlo\ in \qcd, focusing on non-singlet flowed
twist-\two\ operators up to~$n=6$.  The resulting coefficients have already
been employed in the first numerical determinations of the moments of the
valence pion \pdf\ on ensembles generated by the OpenLat
initiative~\cite{Cuteri:2022erk,Cuteri:2022oms,
  Francis:2022hyr,Francis:2023gcm}.  Preliminary results were presented in
\citere{Francis:2024koo}, and the complete analyses are reported in
\citeres{Francis:2025pgf,Francis:2025rya}, which use the results of the
present \nnlo\ calculation to perform the matching to the \msbar\ scheme.

The remainder of this paper is structured as follows.  The theoretical
framework is outlined in
\cref{sec:theoretical_framework}, the main results are presented in
\cref{sec:results}, and conclusions are given in
\cref{sec:conclusions}. The appendix collects the essentials of the \gff, as
well as a description of the contents of the ancillary files associated with
this paper.


\section{Theoretical framework}
\label{sec:theoretical_framework}


\subsection{Twist-two non-singlet operators}\label{sec:ops}

The quantities of interest in this work are the twist-\two\ quark bilinear
operators\footnote{We adopt the same notation for these operators as
in \citeres{Francis:2025rya,Francis:2025pgf} for consistency.}
\begin{equation}
  \begin{aligned}
    \widehat{O}^{rs}_{\{\mu_1\cdots\mu_n\}}(x)
    = \bar{\psi}^r(x)\,\gamma_{\{\mu_1}
    \lrD_{\mu_2}(x)\cdots
    \lrD_{\mu_n\}}(x)\,\psi^s(x)
    - \text{traces}\,,
    \label{eq:ops:felt}
  \end{aligned}
\end{equation}
built from quarks of flavors $r$ and $s$.  Here, $\lrD_\mu = (D_\mu -
\lD_\mu)/2$ is the symmetrized covariant derivative, with the right- and
left-acting derivatives $D_\mu$ and $\lD_\mu$.  The braces indicate normalized
symmetrization over Lorentz indices, and the trace subtraction enforces
$\delta_{\alpha\beta}\widehat{O}_{\mu_1\cdots\mu_n}=0$ for any distinct
$\alpha,\beta\in\{\mu_1,\ldots,\mu_n\}$.  In the continuum, these operators
transform irreducibly under the Euclidean rotation group $O(4)$ and therefore
renormalize multiplicatively.  In the calculation of the matching coefficients
we work with massless quarks and off-diagonal operators $\widehat{O}^{rs}$
with $r\neq s$, which correspond to inserting a particular traceless SU($\nf$)
generator, where $\nf$ is the number of quark flavors.  Since all non-singlet
flavor structures, including phenomenologically relevant combinations such as
$\widehat{O}^{uu}-\widehat{O}^{dd}$ or
$\widehat{O}^{uu}+\widehat{O}^{dd}-2\widehat{O}^{ss}$, are obtained by choosing
other traceless generators and have the same renormalization pattern, the
resulting matching coefficients apply universally to every non-singlet
operator.

For later use we also introduce their Fourier transform,
\be
\widehat{O}(q) = \int\dd^4y\,e^{iqy}\,\widehat{O}(y)\,,
\label{eq:ops:ipoh}
\ee
with $\widehat{O}\equiv\widehat{O}(q{=}0)$ for the corresponding local operator.

For a hadron $h(p)$ with four-momentum $p$, the renormalized matrix elements
of these operators determine the Mellin moments of the \pdf{}s
\be
Z_n\braket{h(p)|\widehat{O}^{qq}_{\{\mu_1\cdots\mu_n\}}|h(p)}(\mu)
= 2 \big(p_{\mu_1}\cdots p_{\mu_n} - \text{traces}\big)\,
  \langle x^{\,n-1}\rangle^{h}_q(\mu)\,,
\label{eq:relmom}
\ee
where the renormalization constants $Z_n$ are collected in \cref{app:ren} up
to $n=6$ in the \msbar\ scheme, and $\mu$ is the renormalization scale.
The $(n{-}1)^\text{th}$ Mellin moment is defined as
\be
\langle x^{\,n-1}\rangle^{h}_q(\mu)
= \int_0^1 \dd x\, x^{n-1}
  \big[q^h(x,\mu) + (-1)^n \qbar^h(x,\mu)\big]\,,
\label{eq:ops:frow}
\ee
where $q^{h}(x,\mu)$ and $\qbar^{h}(x,\mu)$ are the quark and antiquark
distributions in the hadron $h$ at factorization scale $\mu$, respectively.
This
establishes a direct connection between \pdf\ moments and matrix elements of
local renormalized operators.  A lattice calculation of
$\braket{h(p)|\widehat{O}^{qq}_{\{\mu_1\cdots\mu_n\}}|h(p)}$ therefore
provides non-perturbative access to $\langle x^{n-1}\rangle^{h}_q$.

Progress in this direction has long been limited by the loss of full
rotational symmetry once a lattice regulator is introduced.  It causes the
traceless operators in \cref{eq:ops:felt} to mix with operators of lower
dimension~\cite{Kronfeld:1984zv,Martinelli:1987bh,Martinelli:1987zd}, giving
rise to power divergences in the lattice spacing.  For $n\le 4$ this problem
can be partially mitigated by forming specific linear combinations of the
operators $\widehat{O}^{rs}_{\{\mu_1\cdots\mu_n\}}$ that transform according
to irreducible representations of the hypercubic group $\mathrm{H}(4)$.
However, this procedure requires the use of boosted hadrons to project the
correct tensor structures, which leads to a severe degradation of the
signal-to-noise ratio in the relevant correlation functions.  As a
consequence, calculations of $\langle x^{n-1}\rangle$ for $n > 4$ have been
effectively out of reach, and even the moments for $n=3$ and $n=4$ have
traditionally suffered from very large statistical uncertainties.

To solve this problem a strategy has been proposed in
\citere{Shindler:2023xpd}, based on
the \gff~\cite{Narayanan:2006rf,Luscher:2010iy,Luscher:2011bx,Luscher:2013cpa},
which makes it possible to take the continuum limit while restoring rotational
symmetry. This crucially eliminates the power-divergent mixings that affect
the standard lattice operators.  Once the flowed matrix elements are obtained,
the matching to the physical (unflowed) matrix elements can be performed
entirely in the continuum using
the \sftx\ \cite{Luscher:2011bx,Suzuki:2013gza,Luscher:2013vga}.  In this work
we provide the required matching coefficients at \nnlo\ for the non-singlet
flavor sector, covering moments up to $n=6$.


\subsection{The \qcd\ gradient flow}\label{sec:gff}

In the \gff\ the gauge and fermion fields of \qcd\ are extended into an
auxiliary fifth coordinate, the flow time
$t$~\cite{Luscher:2010iy,Luscher:2011bx,Luscher:2013cpa}.  Their evolution in
$t$ is governed by Lorentz-covariant first-order differential equations, with
the boundary condition that at $t=0$ the flowed fields coincide with the
ordinary four-dimensional \qcd\ fields.  Further details on the flow equations
are given in \cref{sec:gradflow}.

A key property of the \gff\ is that composite operators built from flowed
fields at $t>0$ are finite once the standard \qcd\ parameters and the flowed
fermion fields are
renormalized~\cite{Luscher:2011bx,Luscher:2013cpa,Hieda:2016xpq}.  In our case
this requires only the renormalization of the strong coupling (in the \msbar\
scheme) and of the flowed quark fields $\chi^q(t)$.  For the latter we use the
ringed scheme~\cite{Makino:2014taa}, defined by\footnote{Note the factor of
two between the definition of $\overset\leftrightarrow{\mathcal{D}}_\mu$
used here
and in \citere{Artz:2019bpr}.}
\be
\mathring{Z}_\chi\,
\sum_{q=1}^{\nf}\big\langle \chibar^q(t)\,
\overset\leftrightarrow{\slashed{\mathcal{D}}}(t)\,
\chi^q(t)\big\rangle
= -\,\frac{\nc\nf}{(4\pi t)^2}\,,
\label{eq:ringeddef}
\ee
where $\overset\leftrightarrow{\mathcal{D}}_\mu = (\mathcal{D}_\mu
- \overset\leftarrow{\mathcal{D}}_\mu)/2$ is the symmetrized flowed covariant
derivative in the fundamental representation (see \cref{eq:grad:empt}).
The explicit expression for $\mathring{Z}_\chi$ is given
in \cref{eq:jube,eq:gammachi,eq:ringedfin}.

With these ingredients one defines the flowed twist-\two\ operators
\be
\wmcO^{rs}_{\{\mu_1\cdots\mu_n\}}(t,x)
= \mathring{Z}_\chi\,
  \chibar^r(t,x)\,\gamma_{\{\mu_1}
  \overset\leftrightarrow{\mathcal{D}}_{\mu_2}(t,x)\cdots
  \overset\leftrightarrow{\mathcal{D}}_{\mu_n\}}(t,x)\,
  \chi^s(t,x)\,.
\label{eq:ops:flon}
\ee

For sufficiently short flow time, these operators satisfy a \sftx\ relating them to the ordinary twist-\two\ operators of \cref{eq:ops:felt}:
\be
\wmcO^{rs}_{\{\mu_1\cdots\mu_n\}}(t,x)
= \zeta_n^\bare(t)\,
  \widehat{O}_{\{\mu_1\cdots\mu_n\}}(x)\,,
\label{eq:ops:glob}
\ee
which holds up to terms that vanish in the limit $t\to 0$.
For the non-singlet case considered here the matching is multiplicative: each $(n{-}1)^\text{th}$ moment requires a single coefficient $\zeta_n$.

The superscript ``$\bare$'' indicates that these matching coefficients contain ultraviolet divergences associated with the regular operators in \cref{eq:ops:felt}.  
Renormalizing those operators yields the finite coefficients
\be
\zeta_n(t,\mu) = (Z_n\,\zeta_n^\bare(t))(\mu)\,,
\label{eq:ops:fats}
\ee
where $\mu$ is the renormalization scale and $Z_n$ are the renormalization
constants of \cref{eq:relmom} (see \cref{app:ren}).


\subsection{Calculation of the matching coefficients}\label{sec:zetancalc}

For the determination of the matching coefficients $\zeta_n$, we apply the
method of projectors
\cite{Gorishnii:1983su,Gorishnii:1986gn,Harlander:2018zpi}.  Defining
\begin{equation}\label{eq:ops:irma}
  \begin{aligned}
    P_{\mu_1\cdots\mu_n}[O]=N_n
    (\partial_p-\partial_k)^{n-1}_{\mu_2\cdots\mu_n}
    \text{Tr}\left[\gamma_{\mu_1}\braket{\psi_q(p)|O(p+k)|\psi_{q'}(k)}\right]
    \bigg|_{\begin{subarray}{l}
        k=p=0\\
        m_q=m_{q'}=0\end{subarray}} \,,
  \end{aligned}
\end{equation}
where the derivatives $\partial_{p}$, $\partial_{k}$ \wrt\ the quark momenta
and the subsequent nullification of these momenta and the quark masses $m_q$,
$m_{q'}$ are taken \textit{before} any loop integration.  With suitably chosen
normalization factors $N_n$, it follows that
\begin{equation}
  \label{eq:proj}
  P_{\mu_1\cdots\mu_n}[\widehat{O}_{\{\mu_1\cdots\mu_m\}}] = \delta_{nm}
\end{equation}
at tree-level, which can easily be verified using the Feynman rules of the
operators in \cref{eq:ops:felt}.\footnote{In practice, we generate the Feynman
rules as well as the corresponding projectors automatically using in-house
software.}  Up to $n=6$, it is
\begin{equation}
  \begin{aligned}
    \label{normproj}
    N_1&=-\frac{1}{4D}\,,\\
    N_2&=-\frac{i}{2(D+2)(D-1)}\,,\\
    N_3&=-\frac{3}{4(D+4)D(D-1)}\,,\\
    N_4&=-\frac{i}{(D+6)(D+1)D(D-1)}\,,\\
    N_5&=-\frac{5}{4(D+8)(D+2)(D+1)D(D-1)}\,,\\
    N_6&=-\frac{3i}{2(D+10)(D+3)(D+2)(D+1)D(D-1)}\,,
  \end{aligned}
\end{equation} 
where
\begin{equation}\label{eq:ops:bunt}
  \begin{aligned}
    D=4-2\ep
  \end{aligned}
\end{equation}
is the space-time dimension used for regularizing the momentum
integrals\,\cite{tHooft:1972tcz}.  At higher orders, all occurring loop
integrals on the \lhs\ of \cref{eq:proj} are scaleless and thus vanish in
dimensional regularization, which means that this relation actually holds to
all orders in perturbation theory.

Applying the projectors of \cref{eq:proj} onto \cref{eq:ops:glob} therefore
gives the bare matching coefficients~\cite{Harlander:2018zpi,Artz:2019bpr}
\begin{equation}\label{eq:calcmix}
  \zeta^\bare_n(t)=P_{\mu_1\cdots\mu_n}[\wmcO_{\{\mu_1\cdots\mu_n\}}(t)]\,.
\end{equation}
It is important to note that we only included physical operators on the
\rhs\ of \cref{eq:ops:glob}, as they are the only ones that contribute to
physical matrix elements of that relation. For off-shell matrix elements as
required for the method of projectors, unphysical operators have to be taken
into account, in particular those involving total derivatives.  The choice of
the projectors in \cref{eq:ops:irma} is such that they implicitly eliminate
total-derivative operators.

\begin{figure}
  \begin{center}
    \begin{tabular}{c}
      \raisebox{0em}{%
        \mbox{%
          \includegraphics[%
            clip,width=.35\textwidth]%
                          {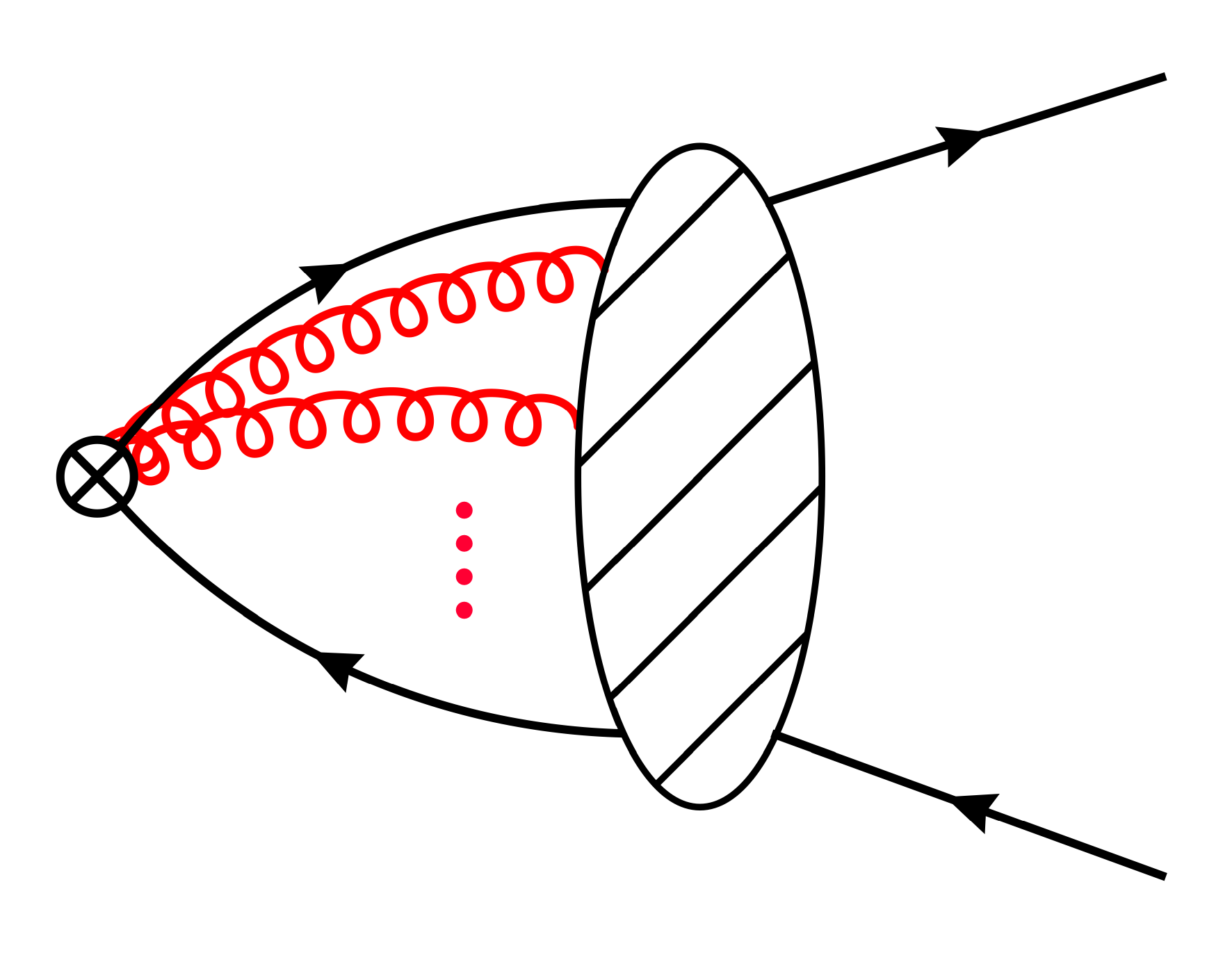}}}
    \end{tabular}
    \parbox{.9\textwidth}{
      \caption[]{\label{fig:generic}\sloppy
        Generic form of the diagrams contributing to the \rhs\ of
        \cref{eq:calcmix}. All Feynman diagrams in this paper were created
        using \code{FeynGame}\,\cite{Harlander:2020cyh,Bundgen:2025utt}.
    }}
  \end{center}
\end{figure}

\begin{figure}
  \begin{center}
    \begin{tabular}{ccc}
      \raisebox{0em}{%
        \includegraphics[width=.23\textwidth]%
                        {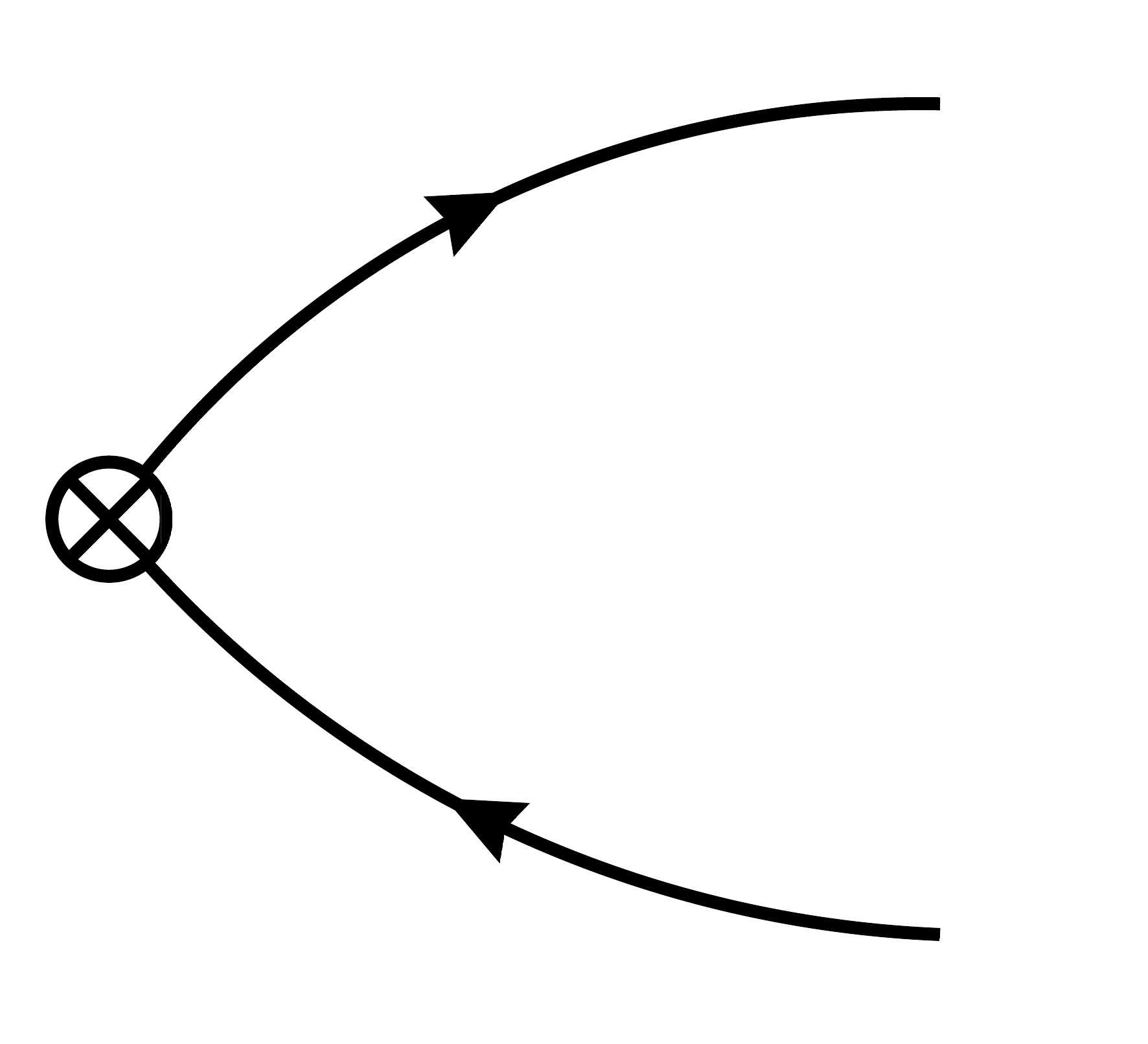}}
      &
      \raisebox{0em}{%
        \includegraphics[width=.23\textwidth]%
                        {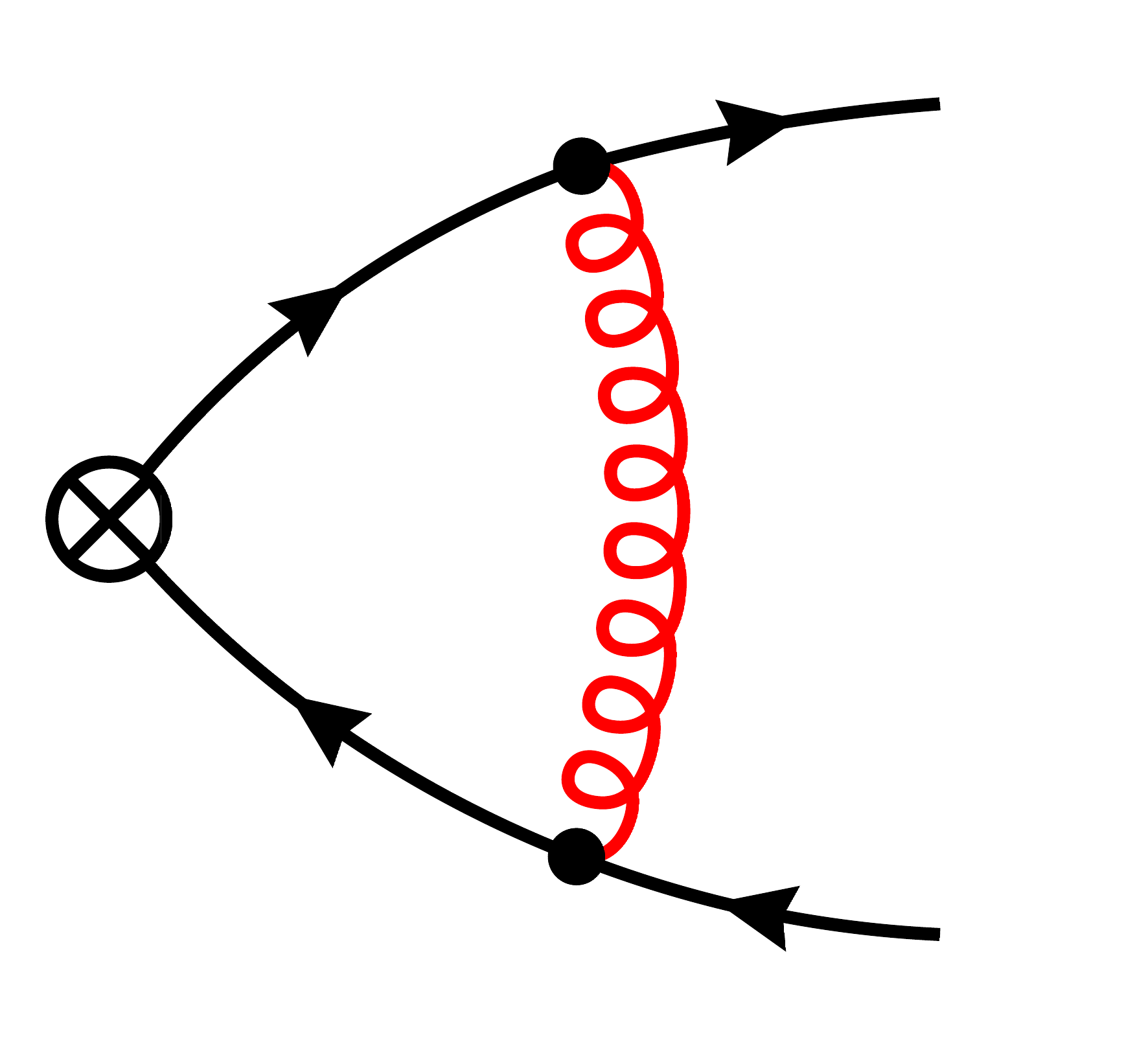}} &
      \raisebox{0em}{%
              \includegraphics[width=.23\textwidth]%
                              {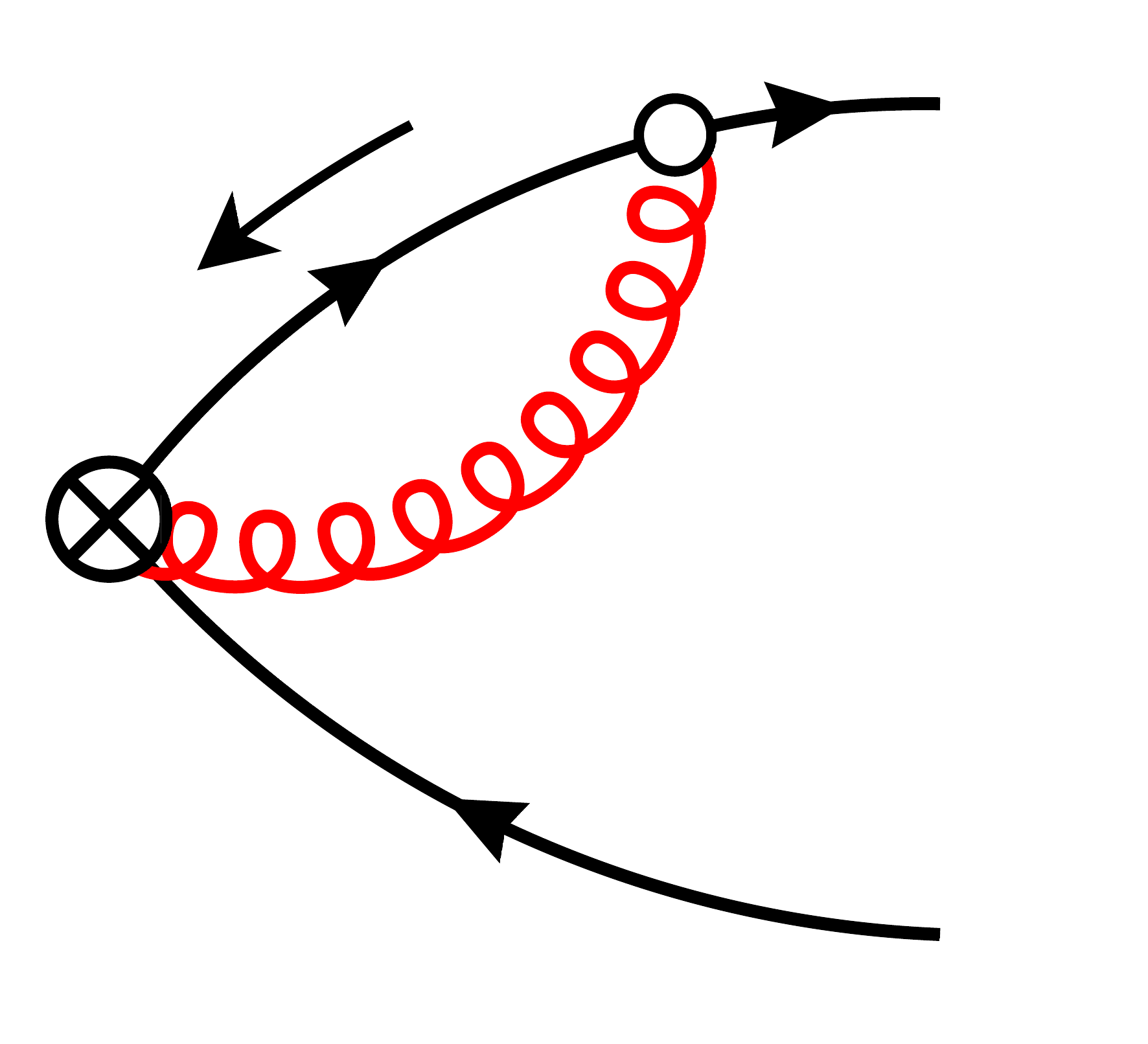}} \\
      (a) & (b) & (c) \\
      \raisebox{0em}{%
              \includegraphics[width=.23\textwidth]%
                              {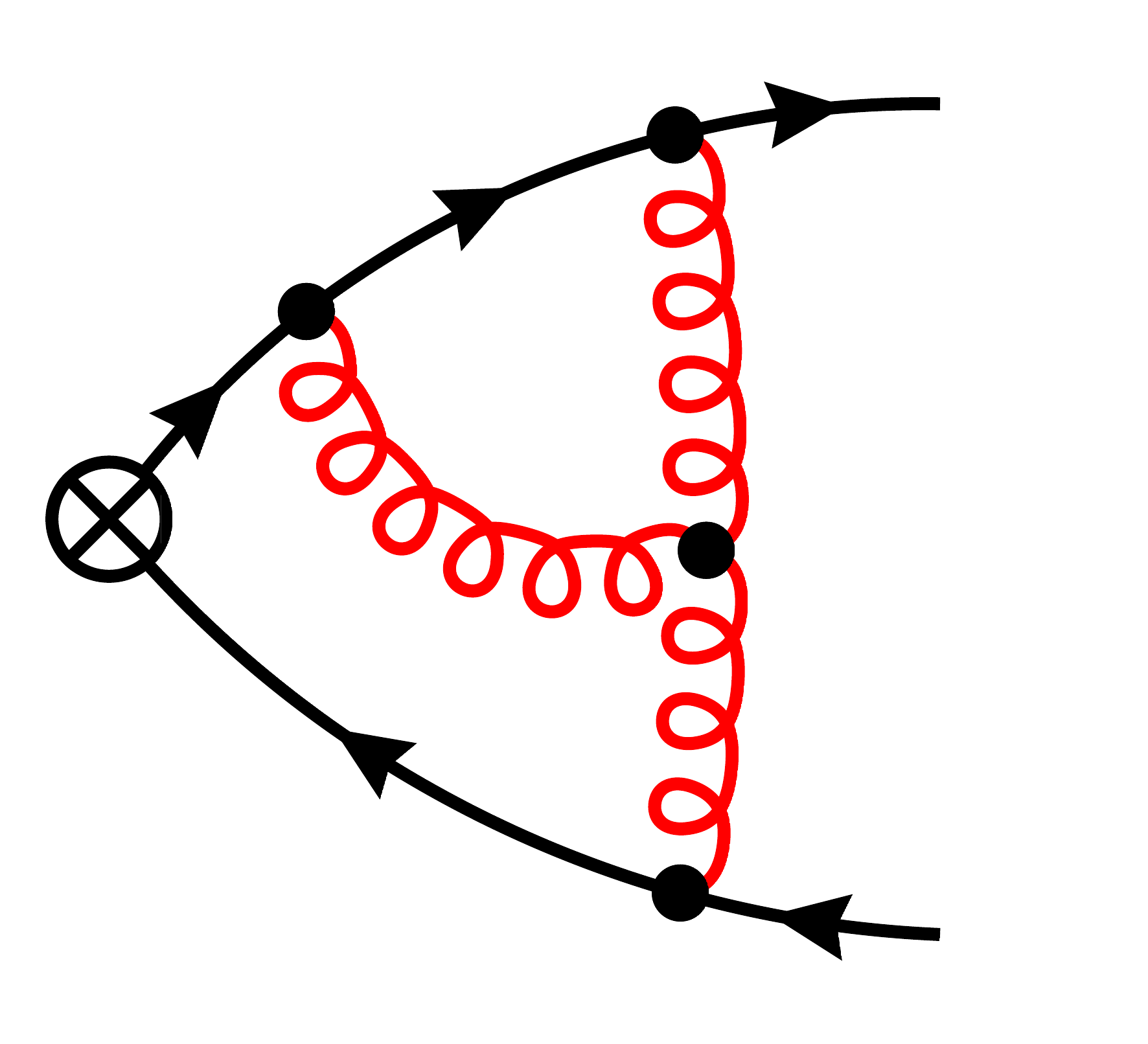}} &
      \raisebox{0em}{%
        \includegraphics[width=.23\textwidth]%
                              {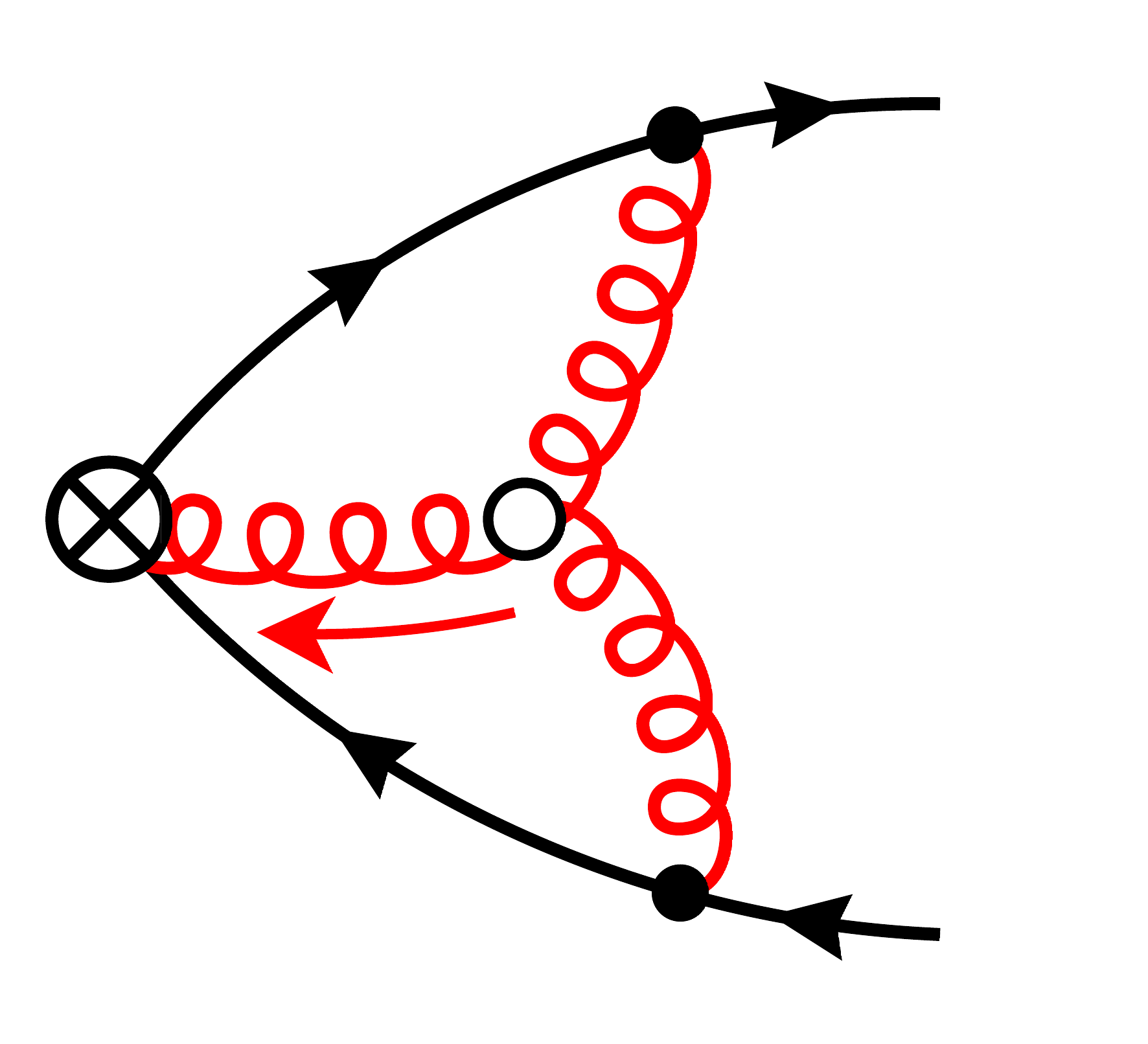}}&
      \raisebox{0em}{%
        \includegraphics[width=.23\textwidth]%
                        {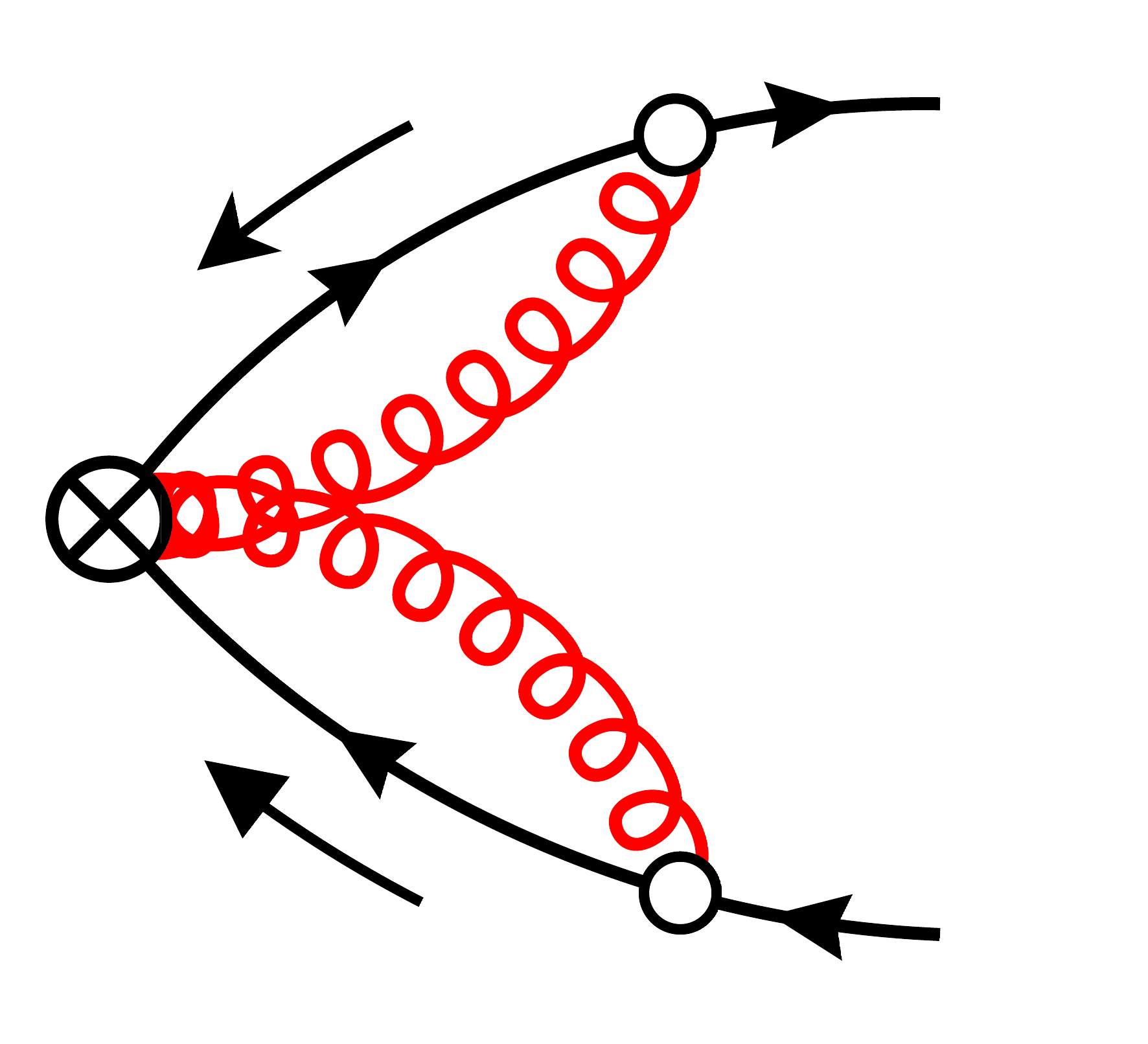}}\\
      (d) & (e) & (f)
    \end{tabular}
    \parbox{.9\textwidth}{%
      \caption[]{\label{fig:dias}\sloppy
        Examples for diagrams contributing to the \rhs\ of
        \cref{eq:calcmix}.
        (a):~tree-level
        diagram contributing for all $n\geq 1$; (b)~and~(c): diagrams
        contributing at \nlo\ for all $n\geq 1$ and $n\geq 2$, respectively;
        (d), (e), (f): diagrams contributing at \nnlo\ for all $n\geq 1,2,3$,
        respectively. Straight lines denote quarks, curly lines gluons; flow
        lines are marked by an arrow next to them, which denotes the flow
        direction; filled/hollow vertex symbols denote regular/flowed
        vertices; the symbol $\otimes$ denotes one of the operators of
        \cref{eq:ops:flon}.  }}
  \end{center}
\end{figure}

The calculation of the matching coefficients according to \cref{eq:calcmix} is
based on an updated version of the setup described in
\citere{Artz:2019bpr}. The Feynman diagrams are generated using
\code{qgraf}~\cite{Nogueira:1991ex,Nogueira:2006pq}.  A generic diagram that
contributes to the \rhs\ of \cref{eq:calcmix} is shown in \cref{fig:generic};
the number of gluons attached to the operator
$\wmcO_{\{\mu_1\cdots\mu_n\}}$ is $\leq n-1$.  Specific examples for
diagrams contributing up to \nnlo\ are shown in \cref{fig:dias}~(a)--(f). The
number of contributing diagrams is shown in \cref{table:ndias}.

Using \code{tapir/exp}~\cite{Harlander:1998cmq,Seidensticker:1999bb,
  Gerlach:2022qnc,Gerlach:2022fgs}, we insert the Feynman rules and identify
the topologies of the associated Feynman integrals. The resulting expression
is evaluated algebraically with the help of
\code{\abbrev{FORM}}~\cite{Vermaseren:2000nd,Ruijl:2017dtg} until it contains
only scalar integrals, multiplied by rational functions depending on the
space-time parameter $D$, see \cref{eq:ops:bunt}. The scalar integrals are
reduced using integration-by-parts relations~\cite{Chetyrkin:1981qh} with the
help of \code{Kira}~\cite{Maierhofer:2017gsa,Klappert:2020nbg} and
\code{FireFly}~\cite{Klappert:2019emp,Klappert:2020aqs}. At \nnlo, this
results in only six (\nlo: one) master integrals which are known
analytically~\cite{Harlander:2018zpi}.


\begin{table}
  \begin{center}
    \caption{\label{table:ndias}
      Number of diagrams contributing to the \rhs\ of \cref{eq:calcmix}
      through \nnlo\ for specific values of $n$.}
    \begin{tabular}{cccccc}
      \hline
      order & $n=1$ & $n=2$ & $n=3$ & $n=4$ & $n\geq 5$\\
      \hline
      \acrshort{LO} & 1 & 1 & 1 & 1 & 1 \\
      \nlo & 10 & 14 & 15 & 15 & 15 \\
      \nnlo & 375 & 737 & 798 & 804 & 805 \\
      \hline
    \end{tabular}
  \end{center}
\end{table}



%
%


%
%


%
\begin{figure}
  \begin{center}
    \begin{tabular}{ccc}
      \raisebox{0em}{%
              \includegraphics[width=.29\textwidth]%
                              {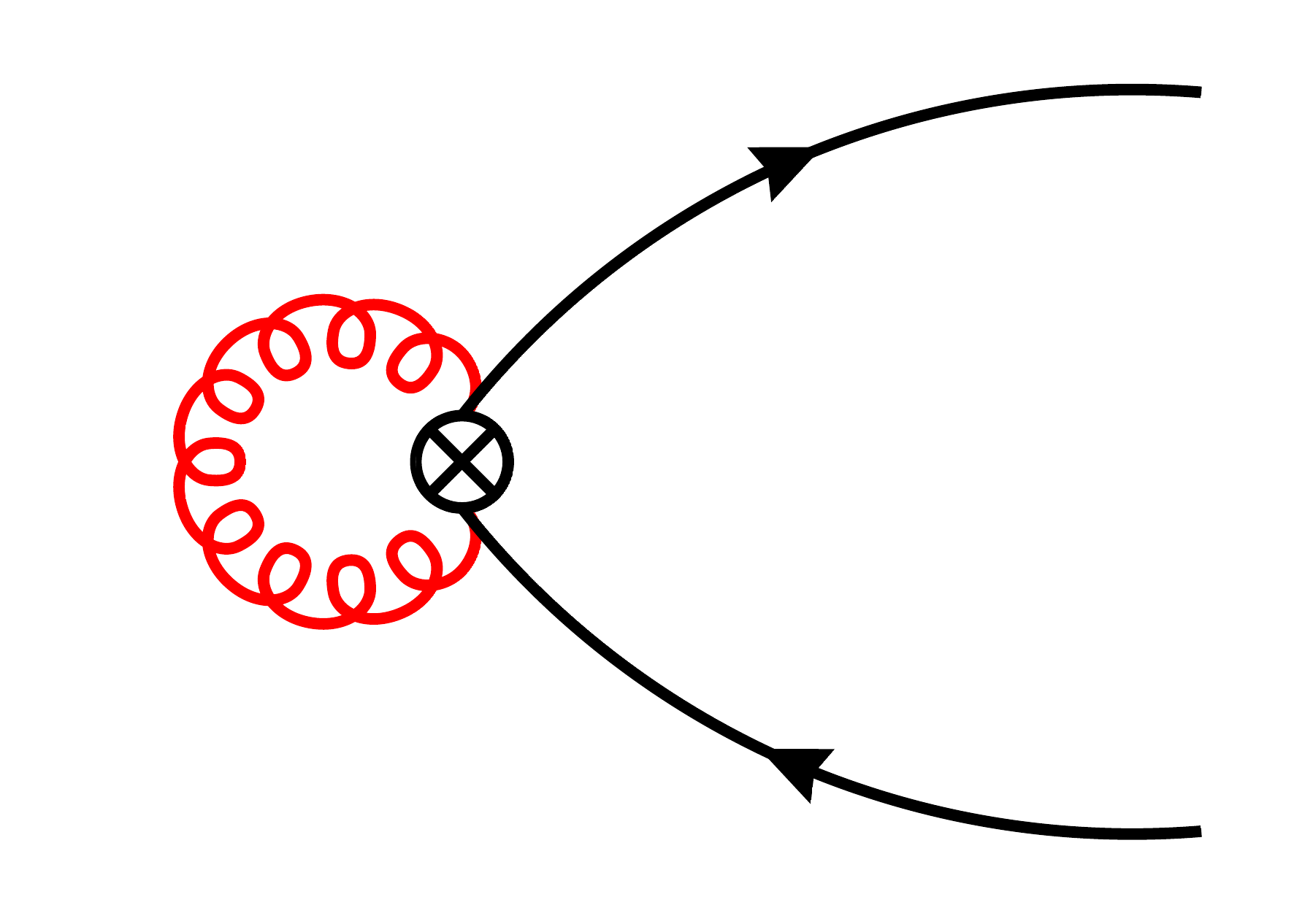}} & 
      \raisebox{0em}{%
              \includegraphics[width=.29\textwidth]%
                              {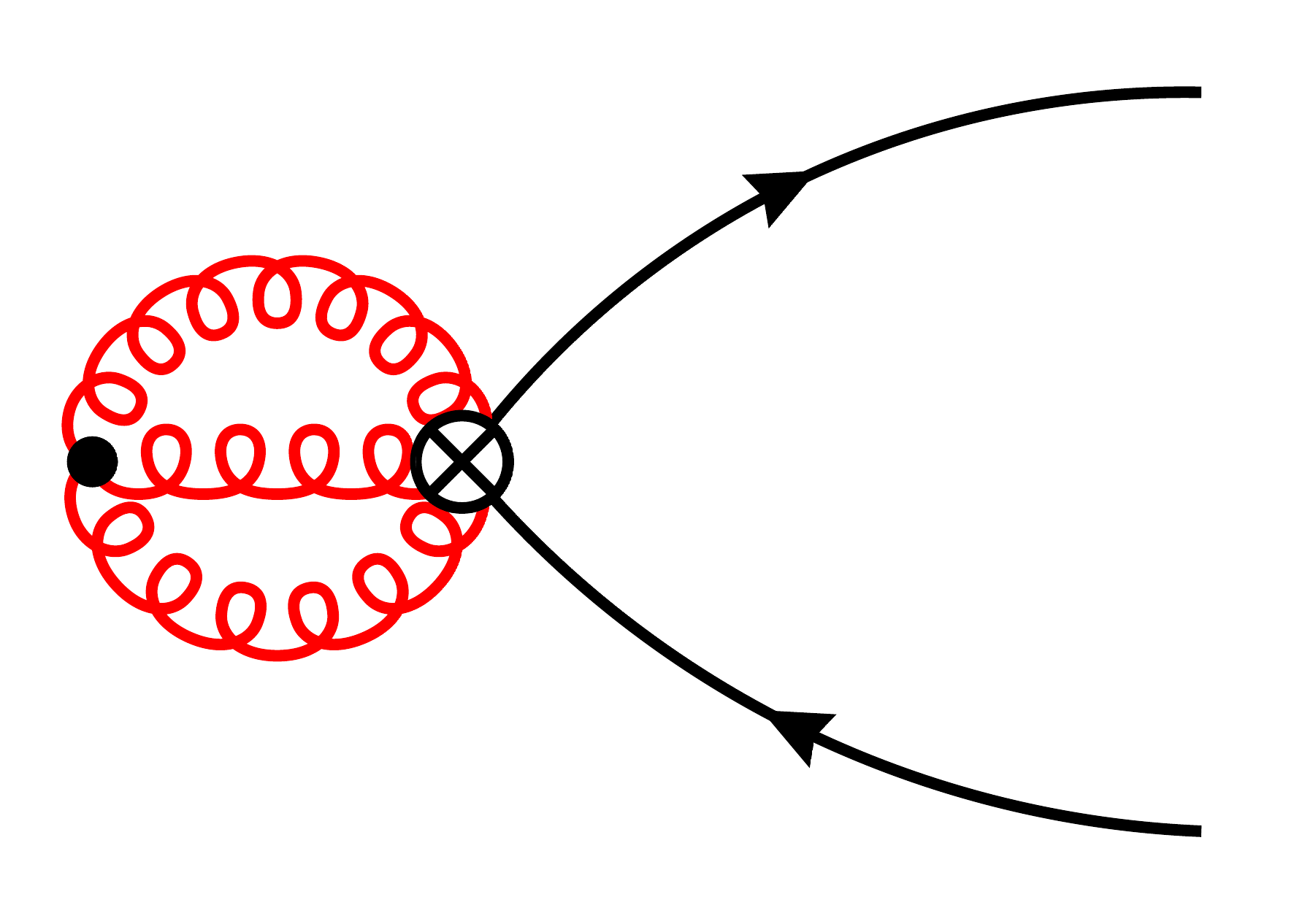}} & 
      \raisebox{0em}{%
              \includegraphics[width=.29\textwidth]%
                              {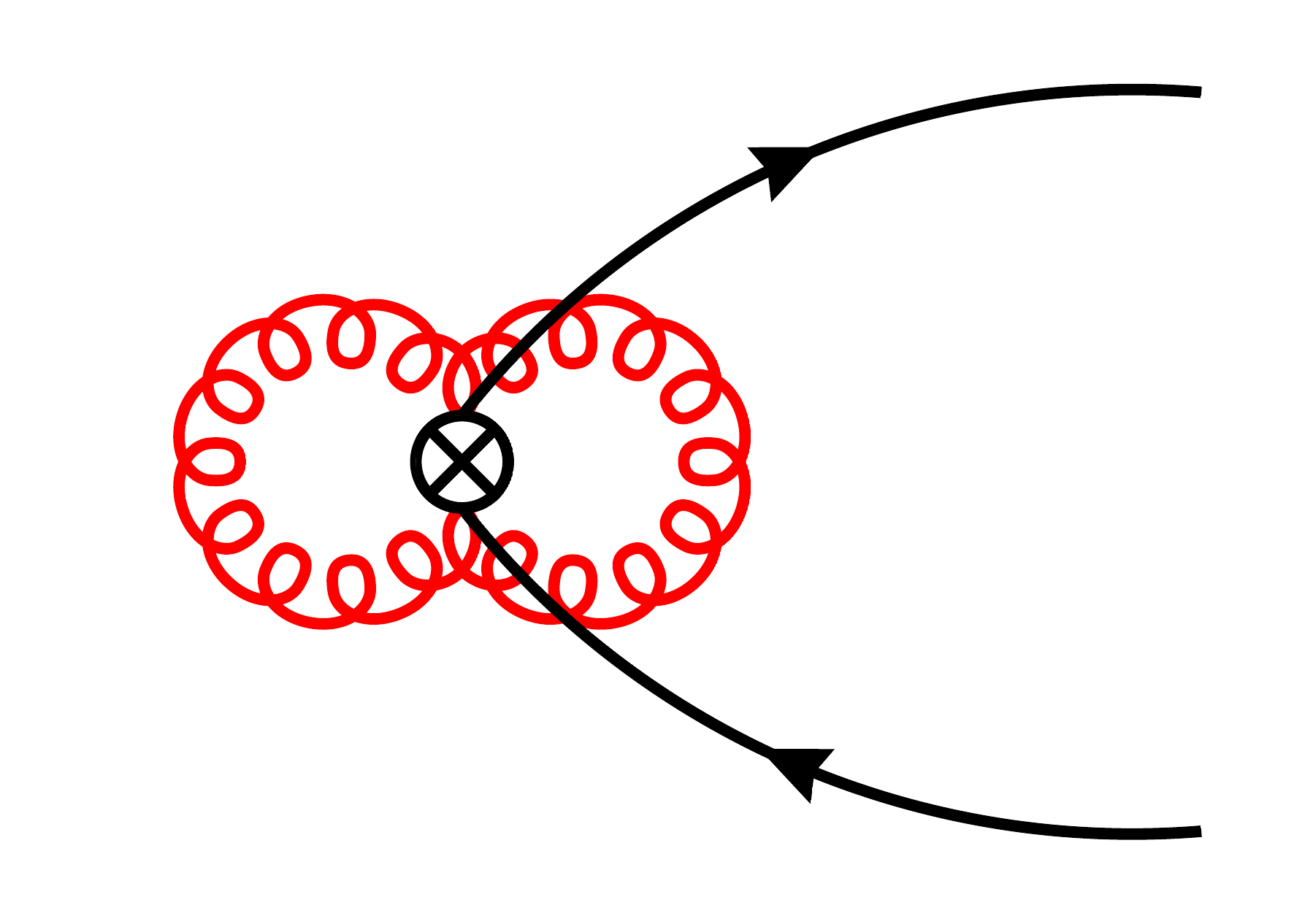}} \\
      (a) & (b) & (c)
    \end{tabular}
    \parbox{.9\textwidth}{%
      \caption[]{\label{fig:bubble}\sloppy Examples for diagrams contributing
        to the \rhs\ of \cref{eq:calcmix}.  (a): diagram contributing at
        \nlo\ for all $n\geq 3$; (b), (c): diagram contributing at \nnlo\ for
        all $n\geq 4,5$, respectively. The notation is the same as in
        \cref{fig:dias}. Note that due to the flow time associated with the
        operator vertex, these diagrams do not involve scaleless integrals and
        are therefore non-zero.}}
  \end{center}
\end{figure}
%

  


\section{Results}\label{}
\label{sec:results}

We provide the results for a general gauge group, with the Casimirs of the
fundamental and adjoint representation denoted by $\ccf$ and $\cca$,
respectively, the trace normalization of the fundamental generators denoted by
$\ctr$, and $\nf$ the number of quark flavors. In \qcd, it is
\begin{equation}\label{eq:kenn}
  \begin{aligned}
    \ccf=\frac{4}{3}\,,\quad \cca=3\,,\quad \ctr=\frac{1}{2}\,.
  \end{aligned}
\end{equation}
Furthermore, we define
\begin{equation}\label{eq:resu:dire}
  \begin{aligned}
    \api(\mu) &= \frac{g^2(\mu)}{4\pi^2}\,,
  \end{aligned}
\end{equation}
where $g(\mu)$ is the strong coupling in the \msbar\ scheme, evaluated
at the renormalization scale $\mu$, see \cref{eq:gren}, and
\begin{equation}\label{eq:resu:elul}
  \begin{aligned}
    \lmut &= \ln 2\mu^2t + \EulerGamma\,,
  \end{aligned}
\end{equation}
where $\EulerGamma = 0.577215\ldots$ is the Euler-Mascheroni constant.
\newcommand\numberthis{\addtocounter{equation}{1}\tag{\theequation}}
Adopting the ringed scheme for the flowed quark fields, we find the
following matching coefficients:
\allowdisplaybreaks
\begin{align*}
    \zeta_1&(t,\mu) =1 + \api\,\ccf\Big(\frac{1}{8} - \ln2
    - \frac{3}{4}\ln3\Big)\\
    & \eqindent{1}
    + \api^2\Bigg\{\frac{1}{16}c_\chi^{(2)} + \ccf^2\Big( - \frac{41}{128}
    - \frac{5}{32}\zeta(2) + \frac{3}{8}\ln2 + \frac{1}{4}\ln^22
    - \frac{3}{32}\ln3 + \frac{3}{2}\text{Li}_2\left(1/4\right)\Big)\\
    & \eqindent{2}
    + \cca\ccf\Big( - \frac{763}{384} - \frac{5}{32}\zeta(2) - \frac{13}{4}\ln2
    + \frac{1}{4}\ln^22 + \frac{27}{8}\ln3
    + \frac{21}{16}\text{Li}_2\left(1/4\right)\Big)\\
    & \eqindent{2}
    + \ccf\ctr\nf\Big(\frac{35}{96} + \frac{1}{8}\zeta(2)\Big)
    + \lmut\bigg[\cca\ccf\Big( - \frac{11}{12}\ln2
      - \frac{11}{16}\ln3 + \frac{11}{96}\Big)
      \\&\eqindent{3}
    + \ccf\ctr\nf\Big(\frac{1}{3}\ln2 + \frac{1}{4}\ln3
    - \frac{1}{24}\Big)\bigg]\Bigg\} + \mathcal{O}(\api^3)\,,
    \numberthis\label{eq:res1}
    \\
    \zeta_2&(t,\mu)=1+\api\ccf\Bigl(\frac{1}{72}-\ln2-\frac{3}{4}\ln3
    +\frac{2}{3}\lmut\Bigr)\\
    &\eqindent{1}+\api^2\Biggl\{\frac{c_\chi^{(2)}}{16}
    +\ccf^2\biggl(\frac{2687}{10368}-\frac{13}{96}\zeta(2)+\frac{865}{216}\ln2
    -\frac{245}{96}\ln3+\frac{1}{4}\ln^22
    +\frac{137}{144}\text{Li}_2\left(1/4\right)\biggr)\\
    &\eqindent{2}+\cca\ccf\biggl(-\frac{12433}{10368}
    -\frac{49}{288}\zeta(2)-\frac{91}{54}\ln2+\frac{7}{6}\ln3+\frac{1}{4}\ln^22
    +\frac{163}{144}\text{Li}_2\left(1/4\right)\biggr)\\
    &\eqindent{2}+\ccf\nf\ctr\biggl(\frac{401}{2592}
    +\frac{1}{72}\zeta(2)\biggr)\\
    &\eqindent{2}+\lmut\biggl[\ccf^2\Bigl(-\frac{1}{4}-\frac{2}{3}\ln2
      -\frac{1}{2}\ln3\Bigr)+\cca\ccf\Bigl(\frac{763}{864}-\frac{11}{12}\ln2
      -\frac{11}{16}\ln3\Bigr)\\&\eqindent{3}+\ccf\nf\ctr\Bigl(-\frac{65}{216}
      +\frac{1}{3}\ln2+\frac{1}{4}\ln3\Bigr)\biggr]\\
    &\eqindent{2}+\lmut^2\Bigl[\frac{2}{9}\ccf^2+\frac{11}{36}\cca\ccf
      -\frac{1}{9}\ccf\nf\ctr\Bigr]\Biggr\}+\mathcal{O}(\api^3)\,,
    \numberthis\label{eq:res2}
    \\
  \zeta_3&(t,\mu)=1+\api\ccf\Bigl(-\frac{11}{288}-\ln2-\frac{3}{4}\ln3
  +\frac{25}{24}\lmut\Bigr)\\
  &\eqindent{1}+\api^2\Biggl\{\frac{c_\chi^{(2)}}{16}
  +\ccf^2\biggl(\frac{42023}{165888}-\frac{119}{1152}\zeta(2)
  +\frac{3145}{1152}\ln2-\frac{1393}{768}\ln3+\frac{1}{4}\ln^22\\
  &\eqindent{3}+\frac{119}{144}\text{Li}_2\left(1/4\right)\biggr)+\cca\ccf\biggl(-\frac{1225}{2592}-\frac{227}{1152}\zeta(2)
  +\frac{527}{576}\ln2-\frac{979}{768}\ln3+\frac{1}{4}\ln^22\\
  &\eqindent{3}+\frac{541}{576}\text{Li}_2\left(1/4\right)\biggr)
  +\ccf\nf\ctr\biggl(\frac{559}{20736}
  -\frac{7}{144}\zeta(2)\biggr)\\
  &\eqindent{2}+\lmut\biggl[\ccf^2\Bigl(-\frac{385}{1152}
  -\frac{25}{24}\ln2
    -\frac{25}{32}\ln3\Bigr)+\cca\ccf\Bigl(\frac{4159}{3456}
    -\frac{11}{12}\ln2-\frac{11}{16}\ln3\Bigr)\\&\eqindent{3}
    +\ccf\nf\ctr\Bigl(-\frac{101}{216}+\frac{1}{3}\ln2
    +\frac{1}{4}\ln3\Bigr)\biggr]\\
  &\eqindent{2}+\lmut^2\Bigl[\frac{625}{1152}\ccf^2+\frac{275}{576}\cca\ccf
    -\frac{25}{144}\ccf\nf\ctr\Bigr]\Biggr\}+\mathcal{O}(\api^3)\,,
    \numberthis\label{eq:res3}
  \\
  \zeta_4&(t,\mu)=1+\api\ccf\Bigl(-\frac{551}{7200}-\ln2-\frac{3}{4}\ln3
  +\frac{157}{120}\lmut\Bigr)\\
  &\eqindent{1}+\api^2\Biggl\{\frac{c_\chi^{(2)}}{16}
  +\ccf^2\biggl(\frac{145703293}{725760000}-\frac{2827}{28800}\zeta(2)
  +\frac{313751}{112000}\ln2-\frac{7028017}{4032000}\ln3+\frac{1}{4}\ln^22\\
  &\eqindent{3}+\frac{1997}{2880}\text{Li}_2\left(1/4\right)\biggr)
  +\cca\ccf\biggl(-\frac{6179}{448000}-\frac{1489}{7200}\zeta(2)
  +\frac{905137}{336000}\ln2-\frac{1018799}{336000}\ln3\\
  &\eqindent{3}+\frac{1}{4}\ln^22
  +\frac{10123}{9600}\text{Li}_2\left(1/4\right)\biggr)
  +\ccf\nf\ctr\biggl(-\frac{57059}{864000}
  -\frac{67}{720}\zeta(2)\biggr)\\
  &\eqindent{2}+\lmut\biggl[\ccf^2\biggl(-\frac{37381}{86400}
    -\frac{157}{120}\ln2-\frac{157}{160}\ln3\biggl)
    +\cca\ccf\biggl(\frac{8213}{5760}-\frac{11}{12}\ln2
    -\frac{11}{16}\ln3\biggl)\\&\eqindent{3}+\ccf\nf\ctr\biggl(-\frac{53}{90}
    +\frac{1}{3}\ln2+\frac{1}{4}\ln3\biggr)\biggr]\\
  &\eqindent{2}+\lmut^2\biggl[\frac{24649}{28800}\ccf^2
    +\frac{1727}{2880}\cca\ccf-\frac{157}{720}\ccf\nf\ctr\biggr]\Biggr\}
  +\mathcal{O}(\api^3)\,,
    \numberthis\label{eq:res4}
  \\
  \zeta_5&(t,\mu)=1
  +\api\ccf\Bigl( -\frac{6163}{57600} -\ln2 -\frac{3}{4}\ln3
  +\frac{91}{60}\lmut\Bigl)\\ &\eqindent{1}
  +\api^2\Biggl\{\frac{c_\chi^{(2)}}{16} +\ccf^2\biggl(
  -\frac{208404649}{8709120000} -\frac{127}{1600}\zeta(2)
  +\frac{150325607}{72576000}\ln2 -\frac{8869939}{8064000}\ln3\\ &\eqindent{3}
  +\frac{1}{4}\ln^22
  +\frac{10441}{14400}\text{Li}_2\left(1/4\right)\biggr)
  +\cca\ccf\biggl(\frac{415266493}{870912000} -\frac{2129}{9600}\zeta(2)
  +\frac{160526033}{36288000}\ln2\\ &\eqindent{3}
  -\frac{301797451}{64512000}\ln3 +\frac{1}{4}\ln^22
  +\frac{1563}{1600}\text{Li}_2\left(1/4\right)\biggr) +\ccf\nf\ctr\biggl(
  -\frac{722059}{5184000} -\frac{23}{180}\zeta(2)\biggr)\\ &\eqindent{2}
  +\lmut\biggl[\ccf^2\biggl( -\frac{361949}{691200} -\frac{91}{60}\ln2
    -\frac{91}{80}\ln3\biggl) +\cca\ccf\biggl(\frac{14717}{9216}
    -\frac{11}{12}\ln2 -\frac{11}{16}\ln3\biggl)\\&\eqindent{3}
    +\ccf\nf\ctr\biggl( -\frac{7891}{11520} +\frac{1}{3}\ln2
    +\frac{1}{4}\ln3\biggr)\biggr]\\ &\eqindent{2}
  +\lmut^2\biggl[\frac{8281}{7200}\ccf^2 +\frac{1001}{1440}\cca\ccf
    -\frac{91}{360}\ccf\nf\ctr\biggr]\Biggr\} +\mathcal{O}(\api^3)\,,
  \numberthis\label{eq:res5}
  \\
  \zeta_6&(t,\mu)=1
  +\api\ccf\Bigl( -\frac{372377}{2822400} -\ln2 -\frac{3}{4}\ln3
+\frac{709}{420}\lmut\Bigl)\\ &\eqindent{1}
+\api^2\Biggl\{\frac{c_\chi^{(2)}}{16} +\ccf^2\biggl(
-\frac{1120571951063}{3651056640000} -\frac{5941}{78400}\zeta(2)
+\frac{95875723}{50935500}\ln2 -\frac{24927078637}{34771968000}\ln3\\
&\eqindent{3}+\frac{1}{4}\ln^22
+\frac{190727}{235200}\text{Li}_2\left(1/4\right)\biggr)
+\cca\ccf\biggl(\frac{116671287151}{134120448000}
-\frac{321851}{1411200}\zeta(2)\\
&\eqindent{3}+\frac{177652851}{30184000}\ln2-\frac{424246536023}{69543936000}\ln3 +\frac{1}{4}\ln^22
+\frac{123721}{117600}\text{Li}_2\left(1/4\right)\biggr)\\ &\eqindent{2}+\ccf\nf\ctr\biggl(
-\frac{355534709}{1778112000} -\frac{197}{1260}\zeta(2)\biggr)\\
&\eqindent{1}+\lmut\biggl[\ccf^2\biggl( -\frac{102996011}{169344000} -\frac{709}{420}\ln2
  -\frac{709}{560}\ln3\biggl)\\&\eqindent{3}
   +\cca\ccf\biggl(\frac{8409979}{4838400}
  -\frac{11}{12}\ln2 -\frac{11}{16}\ln3\biggl)
  +\ccf\nf\ctr\biggl( -\frac{925361}{1209600} +\frac{1}{3}\ln2
  +\frac{1}{4}\ln3\biggr)\biggr]\\ &\eqindent{2}
+\lmut^2\biggl[\frac{502681}{352800}\ccf^2 +\frac{7799}{10080}\cca\ccf
  -\frac{709}{2520}\ccf\nf\ctr\biggr]\Biggr\} +\mathcal{O}(\api^3)\,,
  \numberthis\label{eq:res6}
\end{align*}
where $\text{Li}_2(z)$ is the dilogarithm,
$\zeta(2)=\text{Li}_2(1)=\pi^2/6$,
and $\api=\api(\mu)$.  Through \nlo, our results agree with those
of \citere{Shindler:2023xpd}, where expressions for general $n$ were first
presented.  For $n=1$, the operator in \cref{eq:ops:felt} reduces to the
vector current, whose matching coefficient is known through \nnlo\
from \citere{Borgulat:2023xml}.  The case $n=2$ corresponds to the quark
kinetic operator, with the \nnlo\ matching coefficient computed
in \citere{Harlander:2018zpi}.  Both are consistent with our $\zeta_1$ and
$\zeta_2$ quoted above.  As an additional validation of the higher-$n$
matching coefficients, we carried out the calculation in a general $R_\xi$
gauge and verified that the dependence on the gauge parameter cancels in the
final expressions.

Furthermore, the logarithmic terms can be checked by the
\rg\ equation
\begin{equation}\label{eq:resu:illy}
  \begin{aligned}
    \mu^2\dderiv{}{}{\mu^2}\zeta_n(t,\mu) = \gamma_n\zeta_n(t,\mu)\,,
  \end{aligned}
\end{equation}
where the
\begin{equation}\label{eq:resu:bunk}
  \begin{aligned}
    \gamma_n &= -\api\sum_{i\geq 0}\api^i \gamma_{n,i}
  \end{aligned}
\end{equation}
are the anomalous dimensions of the renormalized twist-two operators. The
values of the coefficients $\gamma_{n,i}$ are given in
\cref{eq:anomdimform,tab:gamma}.


\section{Conclusions}
\label{sec:conclusions}

We have presented the matching coefficients required to relate the first six
non-singlet flowed twist-\two\ operators to the \msbar\ scheme
through \nnlo\ \qcd, providing the perturbative input needed to implement the
method proposed in \citere{Shindler:2023xpd}.  The coefficients have already
been used to match and determine higher moments of the pion
\pdf\ from lattice \qcd\ \cite{Francis:2025rya,Francis:2025pgf}. 
Their applicability is not restricted
to the pion: they can be used equally for proton \pdf{}s and other hadronic
systems.  In addition, since moments of off-forward distributions such as
\gpds\ involve non-forward matrix elements of twist-\two\ operators, our results may be relevant for future studies of \gpds\ based on this approach.

Future work in this direction includes the extension to the singlet sector,
whose treatment is technically more involved due to the enlarged operator
basis and mixing patterns.  We do not expect conceptual obstacles, and the
present calculation provides a solid foundation for its determination.


\paragraph{Acknowledgments.}
A.S. thanks the members of OpenLat for an enjoyable collaboration, in
particular Dimitra A. Pefkou, Jangho Kim, and Andr\'e Walker-Loud for their contributions to the calculation of flowed moments.  R.V.H. and J.T.K. acknowledge support by the
Deutsche Forschungsgemeinschaft (\abbrev{DFG}, German Research Foundation)
under grants 460791904 and 396021762 -- \abbrev{TRR}~257 ``Particle Physics
Phenomenology after the Higgs Discovery''.  A.S. acknowledges funding support
from Deutsche Forschungsgemeinschaft (DFG, German Research Foundation) through
grant 513989149, and from the National Science Foundation grant PHY-2209185.


\begin{appendix}


\section{Definition of the gradient flow}\label{sec:gradflow}

Throughout this paper, we work in $D$-dimensional Euclidean space-time with
$D=4-2\ep$. The flowed gluon field $B^a_\mu(t)$ is defined by the flow
equation~\cite{Narayanan:2006rf,Luscher:2010iy}
\begin{equation}
  \begin{split}
    \partial_t B^a_\mu &= \mathcal{D}^{ab}_\nu G^b_{\nu\mu} + \kappa
    \mathcal{D}^{ab}_\mu \partial_\nu B^b_\nu\,,
    \label{eq:flow}
  \end{split}
\end{equation}
and the initial condition
\begin{equation}
  \begin{split}
    B^a_\mu (t=0) = A^a_\mu\,,
    \label{eq:bound}
  \end{split}
\end{equation}
where $A^a_\mu$ is the regular (i.e.\ unflowed) gluon field,
\begin{equation}\label{eq:grad:bari}
  \begin{aligned}
    G^a_{\mu\nu} &= \partial_\mu B^a_\nu - \partial_\nu B^a_\mu +
    g_\bare f^{abc}B_\mu^b B_\nu^c       
  \end{aligned}            
\end{equation}
is the flowed field-strength tensor,
\begin{equation}\label{eq:grad:hide}
  \begin{aligned}
    \mathcal{D}^{ab}_\mu = \delta^{ab}\partial_\mu - g_\bare f^{abc} B^c_\mu
  \end{aligned}
\end{equation}
is the flowed covariant derivative in the adjoint representation, and
$f^{abc}$ are the \abbrev{SU}(3) structure constants. $g_\bare$ is the bare
strong coupling constant, and $\kappa$ is a gauge parameter which we set to
$\kappa=1$ throughout our calculation. Our final results for the matching
coefficients are independent of this choice though.

The flow equations for the quark fields $\chi(t)$ read~\cite{Luscher:2013cpa}
\begin{equation}
  \begin{aligned}
    \partial_t \chi &= \Delta \chi
    +i \kappa\,\partial_\mu B^a_\mu \frac{\lambda^a}{2} \chi\,,
    &&& \partial_t \bar
    \chi &= \bar \chi \overset\leftarrow \Delta -i \kappa\, \bar \chi
    \partial_\mu B^a_\mu \frac{\lambda^a}{2}\,,
    \label{eq:flowf}
  \end{aligned}
\end{equation}
where the $\lambda^a$ are the Gell-Mann matrices, and the initial conditions
are $\chi (t=0)= \psi$, $\bar{\chi} (t=0)= \bar{\psi}$, where $\psi$,
$\bar\psi$ are the regular quark fields. Furthermore,
\begin{equation}
  \begin{split} \Delta\chi
    = \mathcal{D}_\mu \mathcal{D}_\mu\chi \,,\qquad
    \bar\chi\overset\leftarrow{\Delta}
    = \bar\chi\overset\leftarrow{\mathcal{D}}_\mu \overset\leftarrow{\mathcal{D}}_\mu \,,
  \end{split}
\end{equation}
with the flowed covariant derivative in the fundamental representation,
\begin{equation}\label{eq:grad:empt}
  \begin{aligned}
    \mathcal{D}_\mu = \partial_\mu -i g_\bare B^a_\mu \frac{\lambda^a}{2}
    \,,\qquad
  \overset\leftarrow{\mathcal{D}}_\mu =
    \overset\leftarrow{\partial}_\mu +i g_\bare B^a_\mu
    \frac{\lambda^a}{2}
    \,.
  \end{aligned}
\end{equation}
The perturbative approach for the solution of these equations has been worked
out in \citeres{Luscher:2011bx,Luscher:2013cpa}; more details can be found in
\citere{Artz:2019bpr}.


\section{Renormalization constants}
\label{app:ren}

In the \gff, the strong coupling renormalizes in the same way as in regular
\qcd. We adopt the \msbar\ scheme and relate the bare coupling $g_\bare$ to
the renormalized coupling $g(\mu)$ as
\begin{equation}\label{eq:gren}
  \begin{aligned}
    g_\bare&=\left(\frac{\mu^2e^{\EulerGamma}}{4\pi}\right)^{\epsilon/2}
    Z_g(\api(\mu))g(\mu)\,,&&&
    \api(\mu) &= \frac{g^2(\mu)}{4\pi^2}\,,
  \end{aligned}
\end{equation}
with
\begin{equation}\label{eq:geno}
  \begin{aligned}
    Z_g(\api) &= 1 -
    \api\frac{\beta_0}{2\epsilon}  +
    \api^2\left(\frac{3\beta_0^2}{8\epsilon^2}
    -\frac{\beta_1}{4\epsilon}\right)  +
\mathcal{O}(\api^3)\,,
  \end{aligned}
\end{equation}
and
\begin{equation}\label{eq:anom}
  \begin{aligned}
    \beta_0 &= \frac{1}{4}\left(\frac{11}{3}\cca
    -\frac{4}{3}\ctr\nf\right)\,,&&&
    \beta_1 &= \frac{1}{16}\left[\frac{34}{3}\cca^2-\left(4\ccf
       + \frac{20}{3}\cca\right)\ctr\nf\right]\,.
  \end{aligned}
\end{equation}
The parameters $\ccf$, $\cca$, $\ctr$, and $\nf$ are defined in
\cref{sec:results}.

For the other \msbar\ renormalization constants, we use the form
\begin{equation}\label{eq:gtoZ}
  \begin{aligned}
    Z^{\msbar}(\api) &= 1 - \api\,\frac{\gamma_0}{\epsilon}  +
    \api^2\left[\frac{1}{2\epsilon^2}\left(\gamma^2_{0}  +
      \beta_0\gamma_{0}\right)
      -\frac{\gamma_{1}}{2\epsilon}\right]  +
    \mathcal{O}(\api^3)\,,
  \end{aligned}
\end{equation}
with $\beta_0$ given in \cref{eq:anom}, and the coefficients $\gamma_0$,
$\gamma_1$ to be specified case-by-case.

The renormalization constant of the flowed quark fields in the ringed
scheme, defined by \cref{eq:ringeddef}, is non-minimal and can be written as
\begin{equation}\label{eq:jube}
  \begin{aligned}
    \mathring{Z}_\chi &= \zeta_\chi(t,\mu) Z_\chi^\text{\msbar}\,,
  \end{aligned}
\end{equation}
where the \msbar\ part is given by \cref{eq:gtoZ}, with $\gamma_0$ and
$\gamma_1$ replaced by~\cite{Luscher:2013cpa,Harlander:2018zpi}
\begin{equation}
\label{eq:gammachi}
  \begin{aligned}
    \gamma_{\chi,0} &= -\frac{3}{4}\ccf\,,&&& \gamma_{\chi,1} &=
    \left(\frac{1}{2}\ln 2-\frac{223}{96}\right)\cca\ccf
    +\left(\frac{3}{32}+\frac{1}{2}\ln 2\right)\ccf^2
    +\frac{11}{24}\ccf\ctr\nf\,.
  \end{aligned}
\end{equation}
The finite part reads
\begin{equation}
\label{eq:ringedfin}
  \begin{aligned}
    \zeta_\chi(t,\mu) = 1&
    -\api\left(\gamma_{\chi,0}\lmut
    +\frac{3}{4}\ccf\ln 3+\ccf\ln 2\right)\\
    &+\api^2\Bigg\{\frac{\gamma_{\chi,0}}{2}\left(\gamma_{\chi,0}-\beta_0
    \right)\lmut^2
    +\Big[\gamma_{\chi,0}\left(\beta_0
      -\gamma_{\chi,0}\right)\ln3\\
      &+\frac{4}{3}\gamma_{\chi,0}\left(\beta_0
      -\gamma_{\chi,0}\right)\ln 2-\gamma_{\chi,1}\Big]\lmut
    +\frac{c_\chi^{(2)}}{16}\Bigg\}+\mathcal{O}(\api^3)\,,
  \end{aligned}
\end{equation}
with $\lmut$ defined in \cref{eq:resu:elul}, and~\cite{Artz:2019bpr}
\begin{equation}
\label{eq:chi2}
  \begin{aligned}
    c_\chi^{(2)} &= \cca\ccf\, c_{\chi,\text{A}}
    +\ccf^2\, c_{\chi,\text{F}}+\ccf\ctr\nf\, c_{\chi,\text{R}}\,,
  \end{aligned}
\end{equation}
where
\begin{equation}
\label{eq:chi2coeffs}
  \begin{aligned}
    c_{\chi,\text{A}} &= -23.7947,\qquad c_{\chi,\text{F}}= 30.3914,\qquad\\
    c_{\chi,\text{R}} &=
    -\frac{131}{18}+\frac{46}{3}\zeta_2
    +\frac{944}{9}\ln 2+\frac{160}{3}\ln^2 2
    -\frac{172}{3}\ln 3+\frac{104}{3}\ln2 \ln 3\\
    &-\frac{178}{3}\ln^2 3+\frac{8}{3}\text{Li}_2(1/9)
    -\frac{400}{3}\text{Li}_2(1/3)+\frac{112}{3}\text{Li}_2(3/4)
    =-3.92255\ldots\,.
  \end{aligned}
\end{equation}
The \msbar\ renormalization constants $Z_n$ of the twist-\two\ operators are
determined by \cref{eq:gtoZ}, with $\gamma_i$ replaced by the Mellin moments
$\gamma_{n,i}$ of the splitting functions which are known through three loops
for arbitary values of $n$~\cite{Moch:2004pa,Vogt:2004mw}. For the operators
relevant to this paper, we only require the non-singlet \two-loop expressions
up to $n=6$ \cite{Gross:1973zrg,Floratos:1977au,Gonzalez-Arroyo:1979guc},
which we write as
\begin{equation}
\label{eq:anomdimform}
\gamma_{n,0}=\ccf a_{n}\,,
\qquad\gamma_{n,1}=b_n\ccf^2+c_n\ccf\cca+d_n\ccf\nf\ctr\,,
\end{equation}
where the relevant values of the coefficients $a_n,\ldots,d_n$ are given in
\cref{tab:gamma}. The case $n=1$ is not included in \cref{tab:gamma}, because
$\gamma_{1}=0$ to all orders due to vector current conservation.
\renewcommand{\arraystretch}{1.5}
\begin{table}[htb]
\begin{center}
\caption{Analytical values for the coefficents of $\gamma_{n,0}$ and
  $\gamma_{n,1}$ for the values of $n$ relevant to our renormalization assuming
  the parametrization \cref{eq:anomdimform}.}
\label{tab:gamma}
  \begin{tabular}[t]{ c|cccc}
    \hline
  $n$ & $a_n$ & $b_n$ & $c_n$ & $d_n$\\
  \hline
 $2$ & $\frac{2}{3}$ & $-\frac{7}{27}$ & $\frac{47}{54}$ & $-\frac{8}{27}$ \\
 \hline
 $3$ & $\frac{25}{24}$ & $-\frac{2035}{6912}$ & $\frac{535}{432}$ &
 $-\frac{415}{864}$ \\ 
 \hline
 $4$ & $\frac{157}{120}$ & $-\frac{287303}{864000}$ & $\frac{16157}{10800}$ &
 $-\frac{13271}{21600}$ \\ 
 \hline
 $5$ & $\frac{91}{60}$ & $-\frac{2891}{8000}$ & $\frac{73223}{43200}$ &
 $-\frac{7783}{10800}$ \\ 
 \hline
 $6$ & $\frac{709}{420}$ & $-\frac{3173311}{8232000}$ & $\frac{157415}{84672}$
 & $-\frac{428119}{529200}$ \\
 \hline
\end{tabular}
\end{center}
\end{table}


\section{Ancillary file}

\label{app:anc}
We provide the main results of this paper in computer readable form as
ancillary files \texttt{Zeta.m} in \code{Mathematica} and \texttt{Zeta.py}
in \code{Python} format. A list of the provided quantities is given
in \cref{tab:ancillarycontent}. The matching coefficients $\zeta$ are provided
both in the \msbar\ scheme and the ringed scheme for the flowed quark fields.
Switching between the two is possible by setting the variable
\texttt{Xzetachi} to 0 or 1 to get the $\msbar$ or the ringed scheme,
respectively.\\
All variables and functions which are used in the ancillary files are
collected in \cref{tab:ancvars}. The coefficients $c_\chi^{(2)}$ appearing in
the ringed scheme, see \cref{eq:chi2}, are represented by the
symbol \texttt{C2} and the function \texttt{chi2(nf)} in
\texttt{Zeta.m} and \texttt{Zeta.py} file,
respectively. In the former, its value is contained in the replacement
rule \texttt{ReplaceC2}, see \cref{eq:chi2,eq:chi2coeffs}. We also provide the
anomalous dimensions $\gamma_n$ for general $n$ in the notation
of \cref{eq:anomdimform,eq:resu:bunk}~\cite{Gonzalez-Arroyo:1979guc}. For
$n=2,\ldots, 6$, explicit results are given in \cref{tab:gamma}.
\begin{table}[!ht]
  \begin{center}
    \caption{\label{tab:ancillarycontent} The expressions of the ancillary
    files \texttt{Zeta.m} and \texttt{Zeta.py}
    that encode the main results of this paper. The notation of the
    variables should be self-explanatory; more details are given in the
    header of these files.}
    \begin{tabular}{llll}
      \texttt{Zeta.m} & \texttt{Zeta.py} & meaning & reference\\\hline
      \verb$zeta1$ &
      \verb$zeta_1(alpha_s,nf,mu,t,Oa=2)$ &
      $\zeta_1(t,\mu)$ &
      \cref{eq:res1}\\
      $\quad\vdots$ &
      $\hspace*{3em}\vdots$ &
      $\hspace*{1.5em}\vdots$ &
      $\hspace*{1.5em}\vdots$ \\
      \verb$zeta6$ &
      \verb$zeta_6(alpha_s,nf,mu,t,Oa=2)$ &
      $\zeta_6(t,\mu)$ &
      \cref{eq:res1}\\
      \verb$gamma[n_]$ & \verb$gamma(n,alpha_s,nf,Oa=2)$
      & $\gamma_{n}$ &
      \cref{eq:anomdimform,eq:resu:bunk}
    \end{tabular}
  \end{center}
\end{table}

\begin{table}
  \begin{center}
    \caption{\label{tab:ancvars}Notation for the variables in the
      ancillary files.}
    \begin{tabular}{llll}
      \texttt{Zeta.m} & \texttt{Zeta.py} & meaning & reference\\\hline
      \verb$cf$ &\verb$CF$ &  $\ccf$ &  \cref{eq:kenn} \\
      \verb$ca$ &\verb$CA$ &  $\cca$ &  \cref{eq:kenn} \\
      \verb$tr$ &\verb$TR$ &  $\ctr$ &   \cref{eq:kenn}\\
      \verb$Lmut$&\verb$Lmut$ &  $\lmut$ &  \cref{eq:resu:elul}\\
      \verb$as$ &\verb$a_s$ &$\api$ &  \cref{eq:resu:dire}\\
      \verb$alpha_s$ &\verb$alpha_s$ &$\alpha_s$ &  $\pi \api$\\
      \verb$nf$ &\verb$nf$ &  $\nf$ &  \cref{sec:gff}
    \end{tabular}
  \end{center}
\end{table}
\end{appendix}



\bibliography{literatur}



\end{document}